\documentclass[]{mn2e}
\usepackage{graphicx}
\usepackage{epsfig}
\usepackage{amsmath}
\newcommand       \Angstrom     {\,{\rm \AA}}

\newcommand       \K            {\,{\rm K}}

\newcommand     \gtsim  {\lower.5ex\hbox{$\buildrel > \over \sim$}}
\newcommand     \ltsim  {\lower.5ex\hbox{$\buildrel < \over \sim$}}
\newcommand     \simgt  {\lower.5ex\hbox{$\buildrel > \over \sim$}}
\newcommand     \simlt  {\lower.5ex\hbox{$\buildrel < \over \sim$}}

\newcommand       \mum          {\,{\rm \mu m}}

\newcommand       \Teff         {T_{\rm eff}}

\newcommand       \simali       {\sim\,}

\newcommand       \Aratio        {A_{3.4}/A_{3.3}}

\newcommand       \km        {\,{\rm km}}
\newcommand       \mol       {\,{\rm mol}}

\newcommand       \Adfa       {A_{6.85}}
\newcommand       \Idfa       {I_{6.85}}
\newcommand       \Adfb       {A_{7.25}}
\newcommand       \Idfb       {I_{7.25}}
\newcommand       \Acc        {A_{6.2}}
\newcommand       \Icc        {I_{6.2}}
\newcommand       \NC         {N_{\rm C}}
\newcommand       \NCaro      {N_{\rm C,aro}}
\newcommand       \NCali      {N_{\rm C,ali}}

\newcommand       \CabsPAH    {C^{\scriptscriptstyle\rm PAH}_{\rm abs}}
\newcommand       \Iratiomod   {\left(I_{6.85}/I_{6.2}\right)_{\rm mod}}
\newcommand       \Iratioobs    {\left(I_{6.85}/I_{6.2}\right)_{\rm obs}}
\title{On the Aliphatic versus Aromatic Content
       of the Carriers of
       the ``Unidentified'' Infrared Emission Features
       }
\author[Yang, Glaser, Li, \& Zhong]
       {X.J.~Yang$^{1,2}$\thanks{xjyang@xtu.edu.cn},
        R.~Glaser$^{3}$\thanks{glaserr@missouri.edu},
        Aigen Li$^{2}$\thanks{lia@missouri.edu}, and
        J.X.~Zhong$^{1}$\thanks{jxzhong@xtu.edu.cn}\\
        $^1$Department of Physics,
                      Xiangtan University,
                      411105 Xiangtan, Hunan Province, China\\
        $^2$Department of Physics and Astronomy,
             University of Missouri,
             Columbia, MO 65211, USA\\
        $^3$Department of Chemistry,
             University of Missouri,
             Columbia, MO 65211, USA
             }
\begin{document}
\date{Received date  / Accepted date }
\pagerange{\pageref{firstpage}--\pageref{lastpage}} \pubyear{2016}

\maketitle

\label{firstpage}
\begin{abstract}

Although it is generally accepted that
the unidentified infrared emission (UIE) features
at 3.3, 6.2, 7.7, 8.6, and 11.3$\mum$
are characteristic of the stretching and bending vibrations
of aromatic hydrocarbon materials, the exact nature of
their carriers remains unknown:
whether they are free-flying,
predominantly {\it aromatic} gas-phase molecules,
or amorphous solids
with a {\it mixed aromatic/aliphatic} composition
are being debated.
Recently, the 3.3 and 3.4$\mum$ features which are
commonly respectively attributed to aromatic and
aliphatic C--H stretches have been used to place
an upper limit of $\simali$2\%
on the aliphatic fraction of the UIE carriers
(i.e., the number of C atoms in aliphatic chains
to that in aromatic rings).
Here we further explore the aliphatic
versus aromatic content of the UIE carriers by
examining the ratio of the observed intensity of
the 6.2$\mum$ aromatic C--C feature ($\Icc$)
to that of the 6.85$\mum$ aliphatic
C--H deformation feature ($\Idfa$).
To derive the intrinsic oscillator strengths of
the 6.2$\mum$ stretch ($\Acc$)
and the 6.85$\mum$ deformation ($\Adfa$),
we employ density functional theory
to compute the vibrational spectra of
seven methylated polycyclic aromatic hydrocarbon
molecules and their cations.
%with various aliphatic substituents.
%(e.g., methyl-, dimethyl-, ethyl-, propyl-,
%butyl-PAHs, and PAHs with unsaturated alkyl chains).
%
By comparing $\Idfa/\Icc$ with $\Adfa/\Acc$,
we derive the fraction of C atoms
in methyl(ene) aliphatic form to be at most $\simali$10\%,
confirming the earlier finding that
the UIE emitters are predominantly aromatic.
We have also computed the intrinsic strength
of the 7.25$\mum$ feature ($\Adfb$),
another aliphatic C--H deformation band.
We find that $\Adfa$ appreciably exceeds $\Adfb$. 
%With $\Adfa$ appreciably exceeding $\Adfb$,
This explains why the 6.85$\mum$ feature is more 
frequently detected in space
than the 7.25$\mum$ feature.
\end{abstract}

\section{Introduction}\label{sec:intro}
A series of strong and relatively broad infrared (IR)
emission features at 3.3, 6.2, 7.7, 8.6, and 11.3$\mum$
are ubiquitously seen in almost all astronomical objects
with associated gas and dust,
including protoplanetary nebulae,
planetary nebulae,
reflection nebulae,
young stellar objects,
HII regions,
the Galactic IR cirrus,
and external galaxies (see Tielens 2008).
These features are a common characteristic of
the interstellar medium (ISM) of the Milky Way
and nearby galaxies as well as distant galaxies
out to redshifts of $z\simgt4$
(e.g., see Riechers et al.\ 2014).
Since their first detection four decades ago in
two planetary nebulae (NGC\,7027 and BD$+$30$^{\rm o}$3639,
Gillett et al.\ 1973), the carriers of these IR emission
features have remained unidentified. Because of this,
they are collectively known as
the  ``unidentified infrared emission'' (UIE or UIR) bands.
Nevertheless, it is now generally accepted that these features
are characteristic of the stretching
and bending vibrations of some sorts of
aromatic hydrocarbon materials and therefore,
the UIE features are sometimes also referred to as
the  ``aromatic infrared bands'' (AIB).

A large number of candidate materials have been proposed
as carriers of the UIE bands. All of these materials
contain aromatic structures of fused benzene rings.
The major debate lies in the exact structure of
the UIE carriers:
(1) are they free-flying, predominantly {\it aromatic}
{\it gas-phase} macro-molecules
like polycyclic aromatic hydrocarbon (PAH) molecules
(L\'{e}ger \& Puget 1984; Allamandola et al.\ 1985, 1989),
or (2) are they amorphous {\it solids}
with a {\it mixed aromatic/aliphatic} composition
like hydrogenated amorphous carbon (HAC; Jones et al.\ 1990),
quenched carbonaceous composites (QCC; Sakata et al.\ 1990),
coal or kerogen (Papoular et al.\ 1989)?
As originally suggested by Duley \& Williams (1981),
all of these {\it solid} materials share
the basic molecular structure of PAHs
%by containing aromatic benzene rings.
by containing arenes.\footnote{%
   A benzene ring is C$_6$H$_6$.
   If the H atoms are gone, then it is not really
   ``benzene'' anymore. It is an aromatic ring system
   which can be called ``arene''.
   Arene is a hydrocarbon with alternating double
   and single bonds between carbon atoms forming rings.
   }
They also contain aliphatic C--H bonds as well as
other molecular structures often with other elements
besides C and H.
%Compared to PAHs, these materials
%are bulk in size and their aromatic units contain
%only $\simali$2--4 benzene rings.
The ``MAON'' model recently proposed
by Kwok \& Zhang (2011, 2013) also falls
in this category, where MAON stands for
``{\it mixed aromatic/aliphatic organic nanoparticle}''.

Are the UIE carriers aromatic or aliphatic?
A straightforward way to address this question
is to examine the {\it aliphatic fraction}
of the UIE carriers
(i.e., the fraction of C atoms in aliphatic chains).
Aliphatic hydrocarbons have a vibrational
band at 3.4$\mum$ due to the C--H stretching mode
(Pendleton \& Allamandola 2002).
%In some HII regions, reflection nebulae and planetary nebulae
%%(as well as extragalactic regions)
%the UIE band near 3$\mum$ exhibits a rich spectrum:
%the dominant 3.3$\mum$ feature is usually accompanied
%by a weaker feature at 3.4$\mum$
%along with an underlying plateau
%extending out to $\simali$3.6$\mum$.
%In some objects, a series of weaker features
%at 3.46, 3.51, and 3.56$\mum$ are also seen superimposed
%on the plateau, showing a tendency to decrease in strength
%with increasing wavelength
%(see Geballe et al.\ 1985, Jourdain de Muizon et al.\ 1986,
%Joblin et al.\ 1996).
%%
%While the assignment of the 3.3$\mum$ emission feature to
%the aromatic C--H stretch is widely accepted,
%the precise identification of the 3.4$\mum$ feature
%(and the accompanying weak features at 3.46, 3.51, and 3.56$\mum$
%and the broad plateau) remains somewhat controversial.
%%
In many interstellar and circumstellar environments
the 3.3$\mum$ emission feature is indeed often
accompanied by a weaker feature at 3.4$\mum$
(see Li \& Draine 2012 and references therein).
As demonstrated by Li \& Draine (2012)
and Yang et al.\ (2013), one can place an upper limit
of $\simali$2\% on the aliphatic fraction of the emitters
of the UIE features by assigning the 3.4$\mum$ emission
{\it exclusively} to aliphatic C--H
(also see Rouill{\'e} et al.\ 2012,
Steglich et al.\ 2013).
This is indeed an {\it upper limit}
  as the 3.4$\mum$ emission feature could also be
  due to {\it anharmonicity} of the aromatic C--H
  stretch (Barker et al.\ 1987)
  and ``{\it superhydrogenated}'' PAHs
  whose edges contain excess H atoms
  (Bernstein et al.\ 1996, Sandford et al.\ 2013).

In addition to the 3.4$\mum$ C--H stretching mode,
aliphatic hydrocarbon materials also have two C--H
deformation bands at 6.85$\mum$ and 7.25$\mum$.
These two bands have been observed in weak
{\it absorption} in the Galactic diffuse ISM
(Chiar et al.\ 2000).
They are also seen in {\it emission},
%in interstellar and circumstellar UIE sources
with the 6.85$\mum$ feature detected
both in the Milky Way
and in the Large and Small Magellanic Clouds
while the 7.25$\mum$ feature so far mostly seen
in the Magellanic Clouds
(e.g., see Sloan et al.\ 2014).
Their strengths (relative to the nearby 6.2 and 7.7$\mum$
C--C stretching bands) also allow an estimate of
the aliphatic fraction of the UIE carrier.

In this work, we further explore the aliphatic versus
aromatic content of the UIE carriers by examining
the ratio of the observed intensity of the 6.2$\mum$
aromatic C--C feature ($\Icc$)\footnote{%
  The 7.7$\mum$ C--C stretching feature is
  often stronger than the 6.2$\mum$ C--C stretch
  for most of the interstellar UIE sources.
  However, the IR vibrational spectra of
  methylated-PAHs calculated from density function theory
  are too complicated in the 7.7$\mum$ wavelength range
  to allow us to reliably identify the C--C stretching modes
  from which the 7.7$\mum$ feature originates.
  In contrast, the C--C stretching modes
  in the 6.2$\mum$ wavelength range
  are much ``cleaner''.
  }
to that of the 6.85$\mum$ aliphatic
C--H deformation features ($\Idfa$).
At this moment we will not apply
the 7.25$\mum$ emission feature to constrain
the aliphatic fraction of the UIE carriers
since this feature appears much weaker
than the 6.85$\mum$ feature.\footnote{%
  The detection of the 7.25$\mum$ emission feature
  has so far only been reported in
  several protoplanetary nebulae
  and one planetary nebula
  in the Magellanic Clouds
  (see Sloan et al.\ 2014).
  This feature is not seen
  in the Milky Way except in some protoplanetary disks
around young stars
  (e.g., see Sloan et al.\ 2005, Acke et al.\ 2010).
}
Nevertheless, to help understand why
the 6.85$\mum$ feature is more frequently
detected than the 7.25$\mum$ feature,
we will also compute the intrinsic strength
of the 7.25$\mum$ feature and compare it
with the 6.85$\mum$ feature.
To derive the intrinsic oscillator strengths
of the 6.2$\mum$ aromatic C--C stretch ($\Acc$)
and the 6.85$\mum$ aliphatic C--H deformation ($\Adfa$),
we employ density functional theory
%and second-order perturbation theory
to compute the IR vibrational spectra of
seven methylated polycyclic aromatic
hydrocarbon (PAH) molecules
and their cations.
%with various aliphatic substituents.
%(e.g., methyl-, dimethyl-, ethyl-, propyl-,
%butyl-PAHs, and PAHs with unsaturated alkyl chains).
%
The methyl group is taken to represent
the aliphatic component of the UIE carriers.
PAH cations with aliphatic sidegroups
are of particular interest since ionized PAHs
are expected to be dominant or at least as abundant
as neutrals in astronomical environments
in which the 6.2, 7.7, and 8.6$\mum$ bands are observed.
In \S\ref{sec:Method} we briefly describe
the computational methods
and the parent molecules
based on which we derive the band strengths.
We also present in \S\ref{sec:Method}
the computed band-strength ratios
$\Adfa/\Acc$ and $\Adfb/\Acc$.
We summarize in \S\ref{sec:obs} the intensities
of the 6.2$\mum$ aromatic C--C stretch
and the 6.85 and 7.25$\mum$
aliphatic C--H deformations
observed in the Milky Way and Magellanic Clouds.
In \S\ref{sec:A685A62} we estimate
the aliphatic fractions of the UIE carriers
from the observed intensity ratio
($\Idfa/\Icc$).
We discuss and summarize our major results in
\S\ref{sec:discussion} and \S\ref{sec:summary}, respectively.

%\section{$\Adfa/\Acc$: Band-Strength Ratio of
%         the 6.85$\mum$ Aliphatic C--H Deformation Feature
%         to the 6.2$\mum$ Aromatic C--C Feature
%         \label{sec:Aratio}
%         }
%\subsection{Computational Methods and Target Molecules}
%         \label{sec:Method}
%
%
\section{Computing $\Adfa/\Acc$ and $\Adfb/\Acc$}\label{sec:Method}
We use the Gaussian09 software (Frisch et al.\ 2009)
to calculate the IR vibrational spectra for a range of
methylated aromatic molecules and their cations.
%(see Figure~\ref{fig:MonoMethylPAHs}).
%
We employ the hybrid density functional
theoretical method (B3LYP)
at the {\rm 6-311+G$^{\ast\ast}$} level.\footnote{%
   Yang et al.\ (2013) have carried out computations
   with the B3LYP method
   in conjunction with a variety of basis sets:
   {\rm 6-31G$^{\ast}$},
   {\rm 6-31+G$^{\ast}$},
   {\rm 6-311+G$^{\ast}$},
   {\rm 6-311G$^{\ast\ast}$},
   {\rm 6-31+G$^{\ast\ast}$},
   {\rm 6-31++G$^{\ast\ast}$},
   {\rm 6-311+G$^{\ast\ast}$},
   {\rm 6-311++G$^{\ast\ast}$},
   {\rm 6-311+G(3df,3pd)}, and
   {\rm 6-311++G(3df,3pd)}.
   It is found that the results
   computed with the basis sets
   {\rm 6-311+G$^{\ast\ast}$},
   {\rm 6-311++G$^{\ast\ast}$},
   {\rm 6-311+G(3df,3pd)}, and
   {\rm 6-311++G(3df,3pd)}
   essentially reach the convergence limit.
   The {\rm B3LYP/6-311+G$^{\ast\ast}$} method
   presents an excellent compromise between accuracy
   and computational demand.
   Yang et al.\ (2013) have also examined
   the accuracy of the B3LYP method
   by employing second-order M$\o$ller-Plesset
   perturbation theory (MP2)
   which is thought to be more accurate
   in computing band intensities than B3LYP
   (see Cramer et al.\ 2004).
   It is found that the results
   computed from {\rm B3LYP/6-311++G(3df,3pd)}
   closely agree with that from
   {\rm MP2/6-311++G(3df,3pd)}.
   Therefore, in this work we adopt
   the {\rm B3LYP/6-311+G$^{\ast\ast}$} method.
   %to compute all of the molecules
   %shown in Figure~\ref{fig:MonoMethylPAHs}.
   }
%
%We adopt the precomputed vibrational frequency
%scale factor of $\simali$0.9688 for the
%{\rm B3LYP/6-311+G$^{\ast\ast}$} method.\footnote{%
%  \label{ftnt:scalfac}
%  Taken from the {\it Computational Chemistry
%  Comparison and Benchmark Database} (CCVCBC)
%  at {\rm http://cccbdb.nist.gov/vibscalejust.asp}
%  (accessed 06/01/13)
%  }
We adopt the vibrational frequency
scale factor of $\simali$0.9688 for the
{\rm B3LYP/6-311+G$^{\ast\ast}$} method (Borowski 2012).
We have considered
both neutral and singly-ionized
benzene (C$_6$H$_6$),
naphthalene (C$_{10}$H$_8$),
anthracene (C$_{14}$H$_{10}$),
phenanthrene (C$_{14}$H$_{10}$),
pyrene (C$_{16}$H$_{10}$),
perylene (C$_{20}$H$_{12}$),
and coronene (C$_{24}$H$_{12}$),
as well as all of their methyl derivatives.
%(see Figure ~\ref{fig:MonoMethylPAHs}).
%
%Depending on the position of the methyl attachment,
%a specific aromatic molecule can have
%several structural isomers
%(e.g., the three isomers of methyl-pyrene
%have the ${\rm -CH_3}$ group attached to
%the edge position 1, 2, or 4, respectively;
%see Figure~\ref{fig:MonoMethylPAHs}).
%
%The methyl group is taken to represent the aliphatic
%component of the UIE carriers.
We focus on methyl-substituted PAHs
as PAHs with larger side chains
are not as stable against photolytic
dissociation as methyl-PAHs
and are therefore not expected to
be present in the ISM in a large abundance.

\begin{figure*}
 \vspace{-2mm}
  \begin{center}
  \epsfig{file=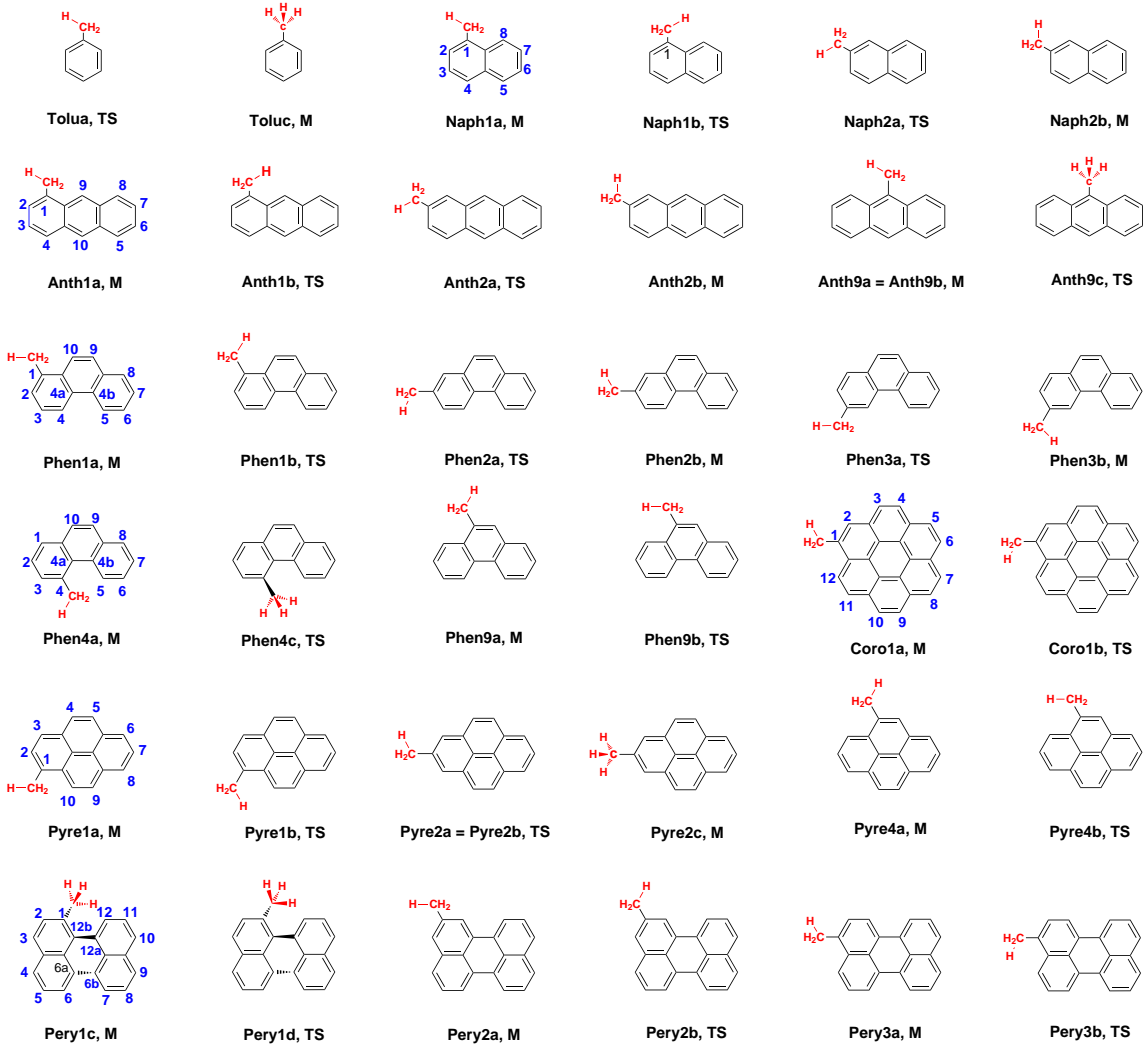,width=15.2cm}
  \end{center}
\vspace{-1mm}
\caption{\label{fig:MonoMethylPAHs}
         Structures of the mono-methyl (${\rm -CH_3}$) derivatives
         of seven aromatic molecules together with the standard IUPAC
         numbering:
         benzene (C$_6$H$_6$),
         naphthalene (C$_{10}$H$_8$),
         anthracene (C$_{14}$H$_{10}$),
         phenanthrene (C$_{14}$H$_{10}$),
         pyrene (C$_{16}$H$_{10}$),
         perylene (C$_{20}$H$_{12}$), and
         coronene (C$_{24}$H$_{12}$).
         We use the first four letters of the parent molecules to
         refer to them and attach the position number of
         the location of the methyl group
         (e.g., Naph1 for 1-methylnaphthalene).
         The mono-methyl derivative of benzene
         is known as toluene (i.e., ``Tolu'', C$_7$H$_8$).
         Depending on where the methyl side-group
         is attached, a molecule will have several
         isomers (e.g., monomethyl-pyrene has three isomers
         in which the -CH$_3$ group is attached
         to carbon 1, 2, or 4, respectively).
         We also indicate whether the structure
         is a minimum (M) or a transition state (TS)
         structure for the methyl rotation.
	 }
\vspace{-3mm}
\end{figure*}

The molecules studied are shown in
Figure~\ref{fig:MonoMethylPAHs}
together with the standard
{\it International Union of Pure and Applied Chemistry}
(IUPAC) numbering scheme.\footnote{%
   http://www.iupac.org
   \label{ftnt:iupac}
   }
We use the first four letters of the molecules
to refer to them and attach the position number
of the location of the methyl group.
For example, 1-methylnaphthalene is referred to as Naph1.
The methyl conformations are indicated
in Figure~\ref{fig:MonoMethylPAHs}
and there are several possibilities.
For their cation counterparts,
we add a ``$+$'' sign
(e.g., Naph1$+$ refers to singly-ionized
1-methylnaphthalene).
%

%%%% Neutral PAHs %%%%

Depending on the symmetry of
the neutral molecule,
there are one or two stereoisomers
in which one of the methyl-CH bonds lies
in the plane of the arene.
We differentiate between these stereoisomers
by addition of ``$a$'' or ``$b$'' to the name of
the structure isomer,
and the in-plane C--H bond points into
the less (more) crowded hemisphere
in the $a$-conformation
($b$-conformation).\footnote{%
   Take Naph1a and Naph1b as examples.
   In Naph1a, the in-plane methyl-H is four bonds
   away from the closest H-atom, H at C2.
   In Naph1b, the in-plane methyl-H is five bonds
   away from the closest H-atom, H at C8.
   This leaves more space between the in-plane
   methyl-H and H(C2) in Naph1a than
   between in-plane methyl-H and H(C8) in Naph1b,
   and Naph1a is less crowded than Naph1b.
   }
In most cases, either the $a$-conformation
or the $b$-conformation corresponds
to the minimum
while the other conformation
corresponds to the transition state structure
for methyl rotation.\footnote{%
    A structure on the potential energy surface is
    a ``stationary structure''
    if the net inter-atomic forces
    on each atom is acceptably close to zero.
    A ``minimum'' is a stationary structure
    for which a small distortion along any
    internal coordinate increases the energy
    (all curvatures are positive).
     A ``transition state structure''
    is a stationary structure for which
    a small distortion along one internal
    coordinate lowers the energy while distortions
    along any of the other internal coordinates
    increases the energy.
    The internal coordinate with the negative curvature
    is called the ``transition vector''.
    For the rotational transition state structures,
    the transition vector describes a rotation of
    the methyl group about the H$_3$C--C bond
    and serves to scramble the H atoms
    in the associated minimum structures
    (i.e., Naph1a can be realized with any
    one of the three methyl-Hs in the plane).
    }
Note that the $a$-conformation can be the minimum (i.e., Naph1a)
or the rotational transition state structure (i.e., Naph2a).

The neutral molecules
Tolu, Anth9 and Pyre2 are symmetric and
the $a$- and $b$-conformations are identical.
In these cases there exists an additional conformation type,
the $c$-conformation, in which one of the methyl-CH bonds is
almost perpendicular with respect to the arene plane.
For Tolu and Pery2, the $c$-conformation is the minimum
while the conformations with in-plane CH-bonds are
the rotational transition state structures.
In contrast, for Anth9 the $c$-conformation
serves as the transition state for interconversion
between the conformations with in-plane CH-bonds.
%

%%%% PAH Cations %%%%

\begin{figure*}
 \vspace{-2mm}
  \begin{center}
  \epsfig{file=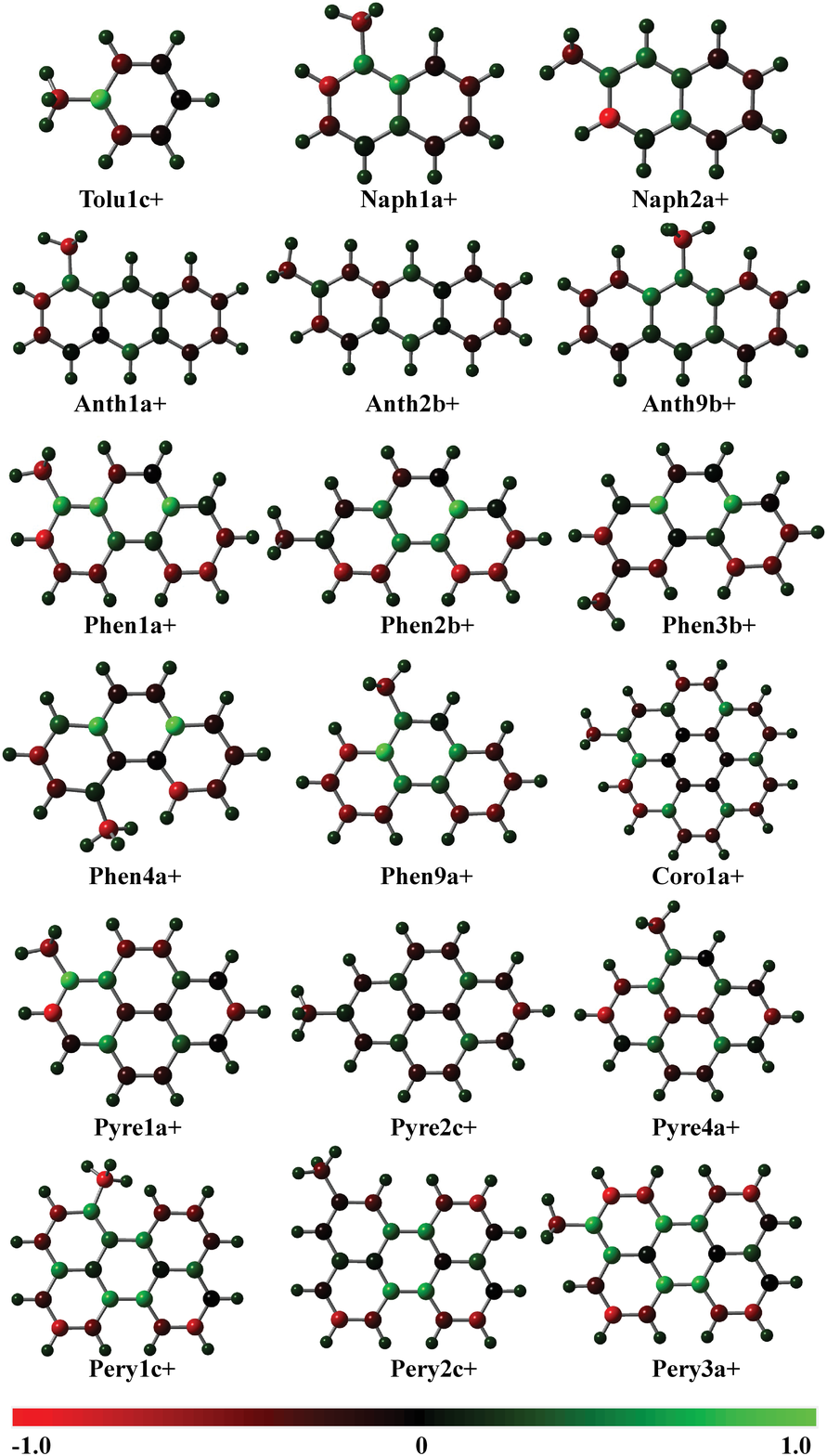,width=10.8cm}
  \end{center}
\vspace{-10mm}
\caption{\label{fig:ChargeDistrib}
         Charge distributions of the cations
         of the mono-methyl (${\rm -CH_3}$) derivatives
         of seven aromatic molecules:
         benzene, naphthalene, anthracene,
         phenanthrene, pyrene, perylene, and coronene.
         We adopt the same naming code as
         in Figure~\ref{fig:MonoMethylPAHs},
         with a ``$+$'' sign
         specifying that the molecule is singly ionized.
         The bottom horizontal bar shows the color coding
         for the charge distribution,
         with red being negatively charged
         and green being positively charged.
         }
\vspace{-3mm}
\end{figure*}

We have also calculated all the minima of
the cations of all the isomers of the methyl
derivatives of the PAH molecules shown in
Figure~\ref{fig:MonoMethylPAHs}.
The structures and Mulliken charge distributions
of these cations are shown in
Figure~\ref{fig:ChargeDistrib}.
The presence of a positive charge
in the $\pi$-system causes polarization
and results in complicated charge patterns
which are characterized by
positively-charged centers and
centers with partial negative charges.
The centers of positive charges are always
tertiary carbons (attached to three C atoms),
while the centers of partial negative charges
are secondary carbons (attached to two C atoms)
and primary carbons (attached to one C atom).
The monocyclic aromatic hydrocarbon Tolu1c+
is atypical in that it is the only cation
with only one tertiary carbon and thus this
carbon is the center of positive charges.
All PAHs contain tertiary carbons at the bonds
shared between rings and these carbons have
the capacity to accept partial positive charges.
Charge delocalization is highly effective
in only a few of the cations (Anth1a+, Anth2b+, Pyre2c+).
Most cations prefer a distribution of charges
over two or more C atoms and concomitant
bond polarizations, with Pery3a+ being an extreme example.

The variety of charge distribution patterns has
several consequences on the vibrational properties.
First, the methyl group may be overall neutral
(i.e., Pyre2c+); it may contain a partially
negative methyl C (i.e., Naph1a+, Antha1a+,
and especially Pery2c+);
the methyl group may be attached to
a cationic center (i.e., Tolu1c+, Phen1a+)
or to an essentially uncharged carbon
(i.e., Anth2b+, Pyre2c+).
As a result, the force constants of
the aliphatic C--H modes will vary much more
in the cations than in the neutral PAHs.
Second, the charge distribution creates
more diversity in the bond strengths of
the aromatic C--H and of the C--C bonds
and the group vibrations will cover broader
spectral ranges.
Third, the presence of a charge
and the resulted charge distribution pattern
affect the dipole polarizabilities and cause
significant differences in the intensities of
the cations as compared to the neutrals.
%
%

%%%% Computational Results %%%%
%
The vibrational frequencies and intensities for
the 6.2$\mum$ aromatic C--C stretching mode
and the 6.85 and 7.85$\mum$ aliphatic
C--H deformation modes of all the isomers of
the seven methylated PAH species
and their cations are computed.
The standard scaling is applied to
the frequencies by employing a scale factor
of $\simali$0.9688. %(see Footnote~\ref{ftnt:scalfac}).
The scaling will not be applied to the intensities
since we are mostly interested in
the intensity ratios $\Adfa/\Acc$ and $\Adfb/\Acc$
and the scaling will be cancelled out.

\begin{figure*}
\centerline{
\includegraphics[scale=0.5,clip]{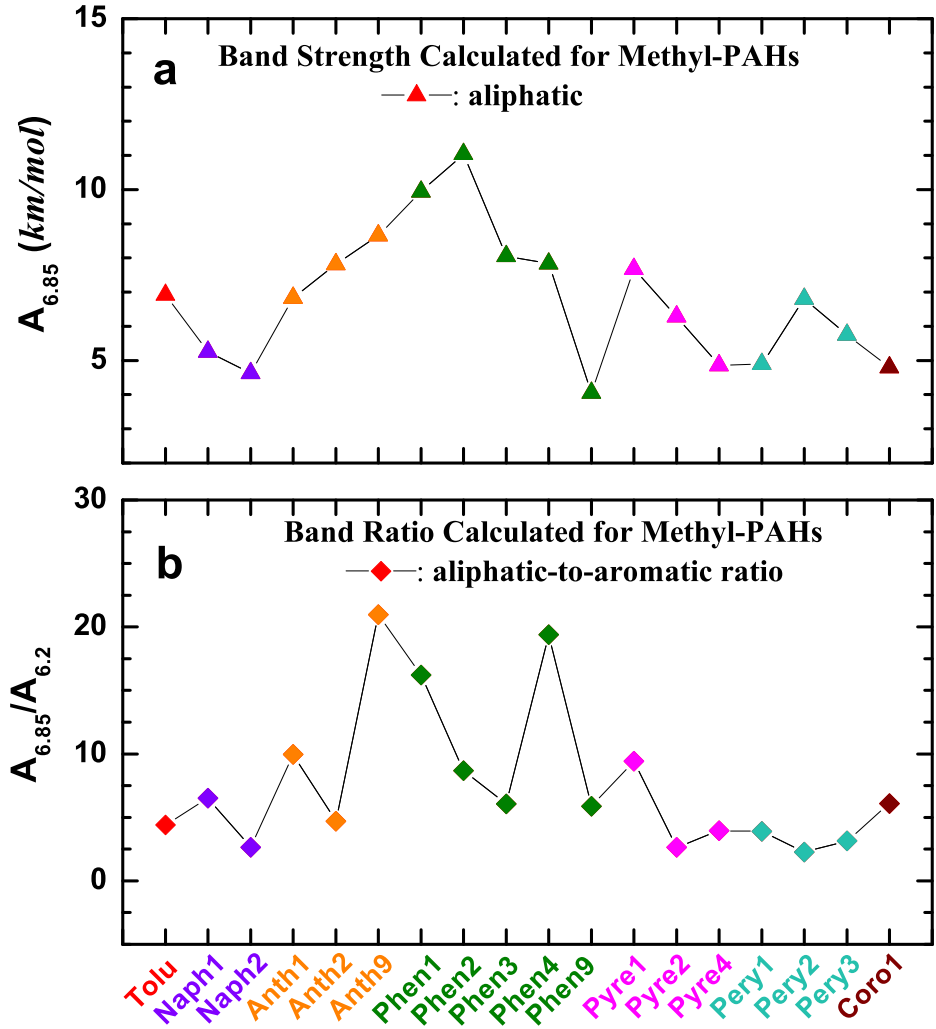}
}
\caption{
         \label{fig:A685A62}
         Band strengths for the 6.85$\mum$ aliphatic
         C--H deformation ($\Adfa$)
         and for the ratio of $\Adfa$ to
         the 6.2$\mum$ aromatic C--C stretch ($\Acc$)
         as determined
         with the B3LYP/{\rm 6-311+G$^{\ast\ast}$} method
         for the mono-methyl derivatives
         of seven aromatic neutral molecules
         and all of their isomers
         (benzene, naphthalene, anthracene,
         phenanthrene, pyrene, perylene, and coronene).
         }
\end{figure*}

\begin{figure*}
\centerline{
\includegraphics[scale=0.5,clip]{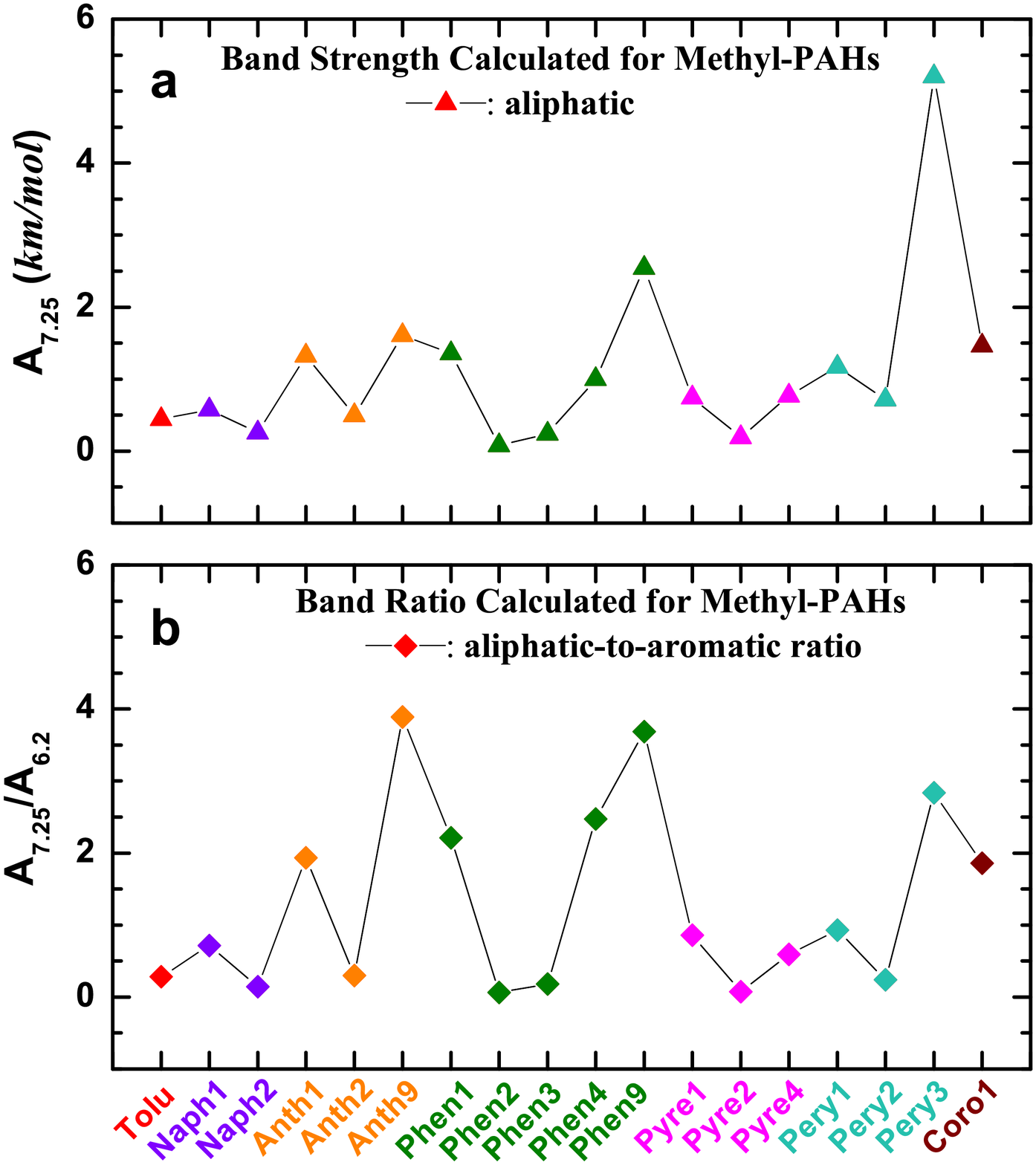}
}
\caption{\label{fig:A725A62}
         Same as Figure~\ref{fig:A685A62}
         but for the 7.25$\mum$ aliphatic
         C--H deformation band ($\Adfb$).
         }
\end{figure*}

We tabulate the computed band strengths
in Table~\ref{tab:A685A62n}
and Table~\ref{tab:A685A62i},
respectively for neutral methyl PAHs
and their cations.
As shown in Figure~\ref{fig:A685A62}a,
both $\Acc$ and $\Adfa$ vary appreciably for
different neutral molecules and their isomers.
For the 6.2$\mum$ aromatic C--C stretch,
it has an average value (per aromatic C-C bond) of
$\langle \Acc\rangle \approx 1.25\km\mol^{-1}$,
with a standard deviation of
$\sigma(\Acc)\approx 0.70\km\mol^{-1}$.
The 6.85$\mum$ aliphatic C--H deformation band
is relatively less dependent on the nature
of the molecule and the specific isomer.
The average band strength (per aliphatic C--H bond)
is $\langle \Adfa\rangle \approx 6.78\km\mol^{-1}$,
and the standard deviation is
$\sigma(\Adfa)\approx 1.93\km\mol^{-1}$.
The $\Adfa/\Acc$ ratio, as shown
in Figure~\ref{fig:A685A62}b,
varies considerably and depends significantly
on the specific molecule and its isomers.
Nevertheless, the $\Adfa/\Acc$ ratio
for neutral molecules clearly
shows a low-end value of $\simali$5
(i.e.,  $\Adfa/\Acc\simgt5$).
For these molecules,
the intrinsic strength of the 7.25$\mum$
aliphatic C--H deformation band, $\Adfb$,
is substantially weaker than the 6.85$\mum$ band.
As shown in Figure~\ref{fig:A725A62}a,
the strength (per aliphatic C--H bond)
of the 7.25$\mum$ band is less than
$\simali$$2\km\mol^{-1}$ for
$\simali$72\% of the molecules
and less than $\simali$$1\km\mol^{-1}$
for $\simali$61\% of the molecules.
Similar to the 6.85$\mum$ band,
although the $\Adfb/\Acc$ ratio varies
considerably for different molecules
and their isomers,
the $\Adfb/\Acc$ ratio also clearly
shows a low-end value of $\simali$1.0
(i.e.,  $\Adfb/\Acc\simgt0.5$;
see Figure~\ref{fig:A725A62}b).
%
%

%%% Results for Cations %%%

\begin{figure*}
\centerline{
\includegraphics[scale=0.5,clip]{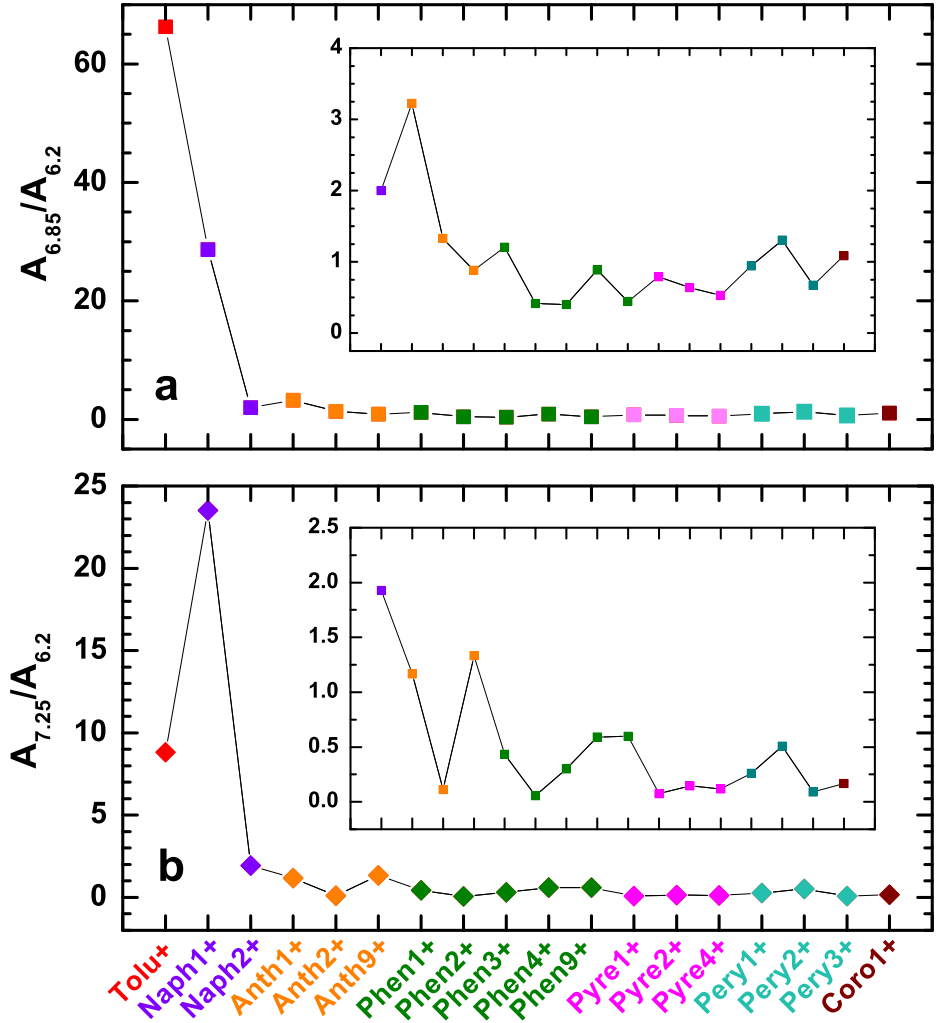}
}
\caption{\label{fig:AratioCation}
         Band-strength ratios of the 6.85 and 7.25$\mum$
         aliphatic C--H deformation bands
         to the 6.2$\mum$ aromatic C--C stretch
         as determined
         with the B3LYP/{\rm 6-311+G$^{\ast\ast}$} method
         for the mono-methyl derivatives
         of seven aromatic cations
         (benzene, naphthalene, anthracene,
         phenanthrene, pyrene, perylene, and coronene)
         and all of their isomers.
         The inset panels enlarge the band-strength ratios
         of those molecules with $\Adfa/\Acc < 4$ (upper panel)
         and with $\Adfb/\Acc < 2$ (bottom panel).
         }
\end{figure*}

For the cations of the methyl PAHs
considered in this work,
the computed band strengths
($\Adfa$, $\Adfb$) and their ratios
($\Adfa/\Acc$, $\Adfb/\Acc$)
are tabulated in Table~\ref{tab:A685A62i}
and shown in Figure~\ref{fig:AratioCation}.
Compared to their neutral counterparts,
the band-strength ratios of cations
show even more substantial differences
among different molecules.
However, similar to neutral molecules,
the band-strength ratios of cations
also show low-end values,
with $\Adfa/\Acc\simgt0.5$
and $\Adfb/\Acc\simgt0.25$
respectively for the 6.85 and 7.25$\mum$ bands.

\section{Observational Constraints}\label{sec:obs}
Unlike the 3.4$\mum$ aliphatic C--H stretch,
the detection of the 6.85 and 7.25$\mum$ aliphatic
C--H deformation bands in the ISM of the Milky Way
is much rarer. We are not aware of any Galactic
sources in which the 7.25$\mum$ feature is seen in
emission except several protoplanetary disks.
On the other hand, to the best of our knowledge,
only seven {\it Galactic} objects exhibit
a noticeable emission feature at 6.85$\mum$:
NGC~7023, a reflection nebula (RN);
NGC~7027, a planetary nebula (PN),
the Orion bar, a photodissociated region (PDR);
and four protoplanetary nebulae
(PPNe; IRAS~01005+7910, IRAS~04296+3429, IRAS~22223+4327,
and IRAS~22272+5435).
Except for the four Galactic PPNe,
the observed 6.85$\mum$ feature intensity
is much weaker than that of the 6.2$\mum$ feature.
As shown in Table~\ref{tab:I685I62},
for Galactic sources other than PPNe, $\Idfa/\Icc\simlt0.10$.
For the four Galactic PPNe,
the 6.85$\mum$ feature is much stronger,
with $\Idfa/\Icc > 1$ for three of these four PPNe.
However, these Galactic PPNe all exhibit
``unusual'' UIE properties
compared to the typical interstellar UIE bands
(e.g., these PPNe show a broad 8.2$\mum$ feature
rather than separately resolved 7.7/8.6$\mum$ UIE features;
moreover, the 11.3 and 12.7$\mum$
C--H out-of-plane bending features
seen in sources with ``normal''-looking UIE features
are blended in a flat-top broad band at $\simali$11.8$\mum$;
see Tokunaga 1997, Peeters et al.\ 2002).
The origin of these  ``unusual'' UIE features is unclear
and their carriers are very likely different from
that of the normal-looking interstellar UIE features.
While the UIE carriers of PPNe are freshly synthesized
in the outflows of asymptotic giant branch (AGB)
and/or post-AGB stars,
their interstellar counterparts must have experienced
various physical and chemical processes in the ISM.
In this work we will restrict ourselves to
the interstellar sources
which show typical-looking UIE features
for which $\Idfa/\Icc\simlt0.10$.

\begin{figure*}
\centerline{
\includegraphics[scale=0.7,clip]{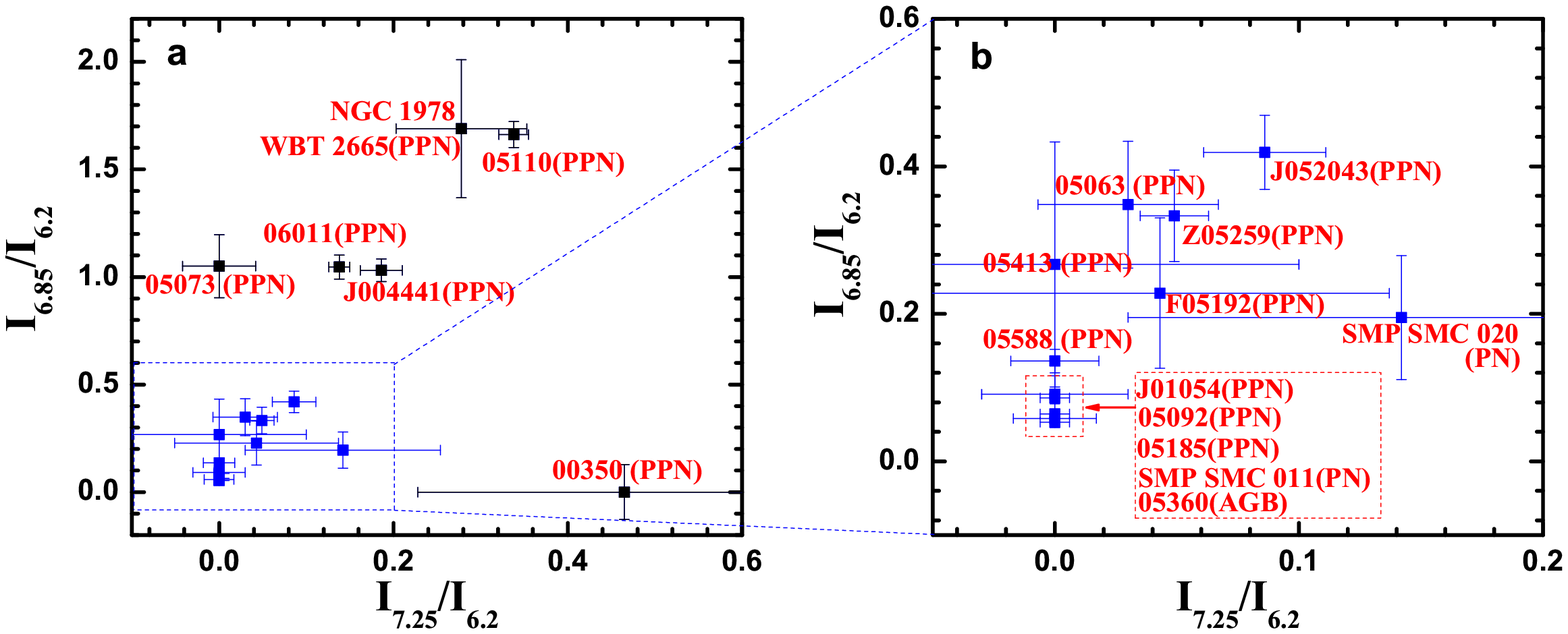}
}
\caption{\label{fig:sloan}
         Observed intensity ratios $\Idfa/\Icc$
         vs. $\Idfb/\Icc$ of the Magellanic Cloud sources
         compiled in Sloan et al.\ (2014).
         The right panel (b) enlarges those sources
         shown in the left panel (a)
         with $\Idfa/\Icc < 0.5$ and $\Idfb/\Icc < 0.2$.
         }
\end{figure*}

It is interesting to note that the 6.85 and 7.25$\mum$
emission features appear to be more frequently
seen in the Magellanic Clouds than in the Galaxy.
Sloan et al.\ (2014)
systematically analyzed the {\it Spitzer}/IRS spectra
of a large number of carbon-rich sources in the Large
and Small Magellanic Clouds. They derived the intensities
of the 6.2, 6.85, 7.25, 7.7 and 8.6$\mum$ emission features
of all the Magellanic sources
which exhibit the 6.85 and 7.25$\mum$ emission features.
We compile the observed intensity ratios
$\Idfa/\Icc$ and $\Idfb/\Icc$
from Sloan et al.\ (2014) for all these sources
and tabulate them in Table~\ref{tab:I685I62}
and show them in Figure~\ref{fig:sloan}.
These sources are predominantly PPNe,
with only three out of eighteen are
not PPNe: SMP\,SMC\,011 (PN), SMP\,SMC\,020 (PN),
and IRAS\,05360-7121 (AGB).
While the Magellanic PPNe have
a wide range of intensity ratios
from $\Idfa/\Icc\simlt0.1$ to $\Idfa/\Icc\simgt1.0$
and from $\Idfb/\Icc\simlt0.05$ to $\Idfb/\Icc\simgt0.2$,
two of the non-PPN sources
(SMP\,SMC\,011, IRAS~05360-7121)
have $\Idfa/\Icc\simlt0.06$
and one non-PPN source (SMP\,SMC\,020)
has $\Idfa/\Icc\simali0.2$.
The 7.25$\mum$ feature is not seen
in SMP\,SMC\,011 and IRAS~05360-7121.
Even in SMP\,SMC\,020,
the only non-PPN source
in the Magellanic Clouds
in which the detection of
the 7.25$\mum$ feature was reported,
the intensity ratio has
a large error bar
(i.e., $\Idfb/\Icc\approx0.14\pm0.11$).
Combining the intensity ratios observed
for both Galactic and Magellanic sources,
in the following we will take
$\Idfa/\Icc = 0.1$ to determine
the aliphatic fractions of the UIE carriers.
The 7.25$\mum$ emission feature will not
be used to estimate the aliphatic content of
the UIE carriers.

\section{Constraints
         on the Aliphatic Fraction of the UIE Carriers
         from the 6.85$\mum$ Features}
         \label{sec:A685A62}

Let $N_{\rm C,ali}$ and $N_{\rm C,aro}$
respectively be the numbers of aliphatic
and aromatic C atoms in the emitters
of the 6--8$\mum$ UIE bands.
Let $B_\lambda\left(T\right)=
\left(2 h c^2/\lambda^{5}\right)
/\left[\exp\left(hc/\lambda kT\right)-1\right]$
be the Planck function at wavelength $\lambda$
and temperature $T$
(where $h$ is the Planck constant,
$c$ is the speed of light, and $k$ is
the Boltzman constant).
%with $B_{6.85}/B_{6.2}\approx1.20\pm0.25$
%for $330<T<1000\K$.
The aliphatic fraction can be estimated from
\begin{equation}
\frac{N_{\rm C,ali}}{N_{\rm C,aro}}
\approx \frac{I_{6.85}}{I_{6.2}}
\times\frac{A_{6.2}}{A_{6.85}}
\times\frac{B_{6.2}}{B_{6.85}}
\times\left(\frac{1.25}{2.5}\right) ~~,
\end{equation}
where the term $\left(1.25/2.5\right)$
arises from that we assume that one aliphatic C atom
corresponds to 2.5 aliphatic C--H bonds
(intermediate between methylene $-$CH$_2$
and methyl $-$CH$_3$)
and one aromatic C atom corresponds
to 1.25 aromatic C--C bond
(intermediate between benzene C$_6$H$_6$
and coronene C$_{24}$H$_{12}$).
The major UIE features fall in the wavelength
range of $\simali$3--12$\mum$,
implying a temperature range of
$200\simlt T\simlt 800\K$.\footnote{%
%  {\bf
  Let $C_{\rm abs}(\lambda)$ be the dust
  absorption cross section at wavelength $\lambda$
  and $j_\lambda$ be the dust IR emissivity.
  For $C_{\rm abs}(\lambda)\propto \lambda^{-\beta}$,
  $\lambda j_\lambda$ peaks at
  $\lambda_p\approx\left(h c/k T\right)/\left(4+\beta\right)$
  (see Li 2005). With $\beta\approx2$ for PAH-like UIE carriers,
  we have $T\approx800\K$ for $\lambda_p\sim3\mum$
  and $T\approx200\K$ for $\lambda_p\sim12\mum$.
%  }
  }
With $I_{6.85}/I_{6.2}\simlt0.10$ and
$B_{6.85}/B_{6.2}\approx1.04\pm0.24$
for $200\simlt T\simlt 800\K$
we obtain an upper limit
of $N_{\rm C,ali}/N_{\rm C,aro}\approx0.96\%$
for neutral methylated-PAHs with $\Acc/\Adfa\simlt0.2$,
and $N_{\rm C,ali}/N_{\rm C,aro}\approx9.6\%$
for ionized methylated-PAHs with $\Acc/\Adfa\simlt2.0$.
We therefore conclude that the carriers of
the 6--8$\mum$ UIE bands are largely aromatic,
with $<$\,10\% of the C atoms in aliphatic form.

The above derivation of $N_{\rm C,ali}/N_{\rm C,aro}$
is based on the assumption that the UIE carriers emit
at a single temperature in the range of
$200\simlt T\simlt 800\K$.
However, the UIE carriers are nano-sized (or smaller)
and will be transiently heated by single stellar photons
(see Li 2004). Illuminated by starlight, they will not
attain an equilibrium temperature, instead, they will
have a distribution of temperatures.
Following Draine \& Li (2001), we will calculate
the temperature probability distribution functions
and emission spectra of methyl PAHs of
$\NCaro$ aromatic C atoms
and $\NCali$ aliphatic C atoms.
For such molecules, we approximate
their absorption cross sections by
adding three Drude functions to that
of PAHs of $\NCaro$ C atoms.
These Drude functions represent
the 3.4$\mum$ aliphatic C--H stretch,
and the 6.85 and 7.25$\mum$ aliphatic
C--H deformations. The absorption cross section
of  a methyl PAH molecule of $\NCaro$ aromatic
C atoms and $\NCali$ aliphatic C atoms becomes
\begin{eqnarray}
C_{\rm abs}(\NC,\lambda) & = & \CabsPAH(\NCaro,\lambda)\\
& + & \NCali \frac{2}{\pi}
    \frac{\gamma_{3.4} \lambda_{3.4} \sigma_{\rm int,3.3}
     \left(A_{3.4}/A_{3.3}\right)}
     {(\lambda/\lambda_{3.4}-\lambda_{3.4}/\lambda)^2
      +\gamma_{3.4}^2}\\
&+& \NCali \frac{2}{\pi}
     \frac{\gamma_{6.85} \lambda_{6.85}
     \sigma_{\rm int,6.2} \left(\Adfa/\Acc\right)}
     {(\lambda/\lambda_{6.85}-\lambda_{6.85}/\lambda)^2
      +\gamma_{6.85}^2}\\
&+& \NCali \frac{2}{\pi}
    \frac{\gamma_{7.25} \lambda_{7.25}
    \sigma_{\rm int,6.2} \left(\Adfb/\Acc\right)}
     {(\lambda/\lambda_{7.25}-\lambda_{7.25}/\lambda)^2
      +\gamma_{7.25}^2} ,
\end{eqnarray}
where $\NC=\NCaro+\NCali$; $\lambda_{3.4}=3.4\mum$,
$\lambda_{6.85}=6.85\mum$, and $\lambda_{7.25}=7.25\mum$
are respectively the peak wavelengths
of the 3.4, 6.85 and 7.25$\mum$ features;
$\gamma_{3.4}\lambda_{3.4}$,
$\gamma_{6.85}\lambda_{6.85}$,
and $\gamma_{7.25}\lambda_{7.25}$
are respectively the FWHMs
of the 3.4, 6.85 and 7.25$\mum$ features
($\gamma_{3.4}$, $\gamma_{6.85}$,
and $\gamma_{7.25}$ are dimentionless parameters;
see Draine \& Li 2007);
and $\sigma_{{\rm int},3.3}$ and $\sigma_{{\rm int},6.2}$
are respectively the integrated strengths per (aromatic)
C atom of the 3.3$\mum$ aromatic C--H stretch
and 6.2$\mum$ aromatic C--C stretch
(see Draine \& Li 2007).
We take $A_{3.4}/A_{3.3}=1.76$ (see Yang et al.\ 2013).
We take the lower limits of $\Adfa/\Acc\approx5.0$
and $\Adfb/\Acc\approx0.5$ for neutrals,
$\Adfa/\Acc\approx0.5$ and $\Adfb/\Acc\approx0.25$
for cations as derived in \S\ref{sec:Method}.
We note that, for a given observed intensity ratio
$\Idfa/\Icc$, a lower limit on $\Adfa/\Acc$ leads
to an upper limit on the aliphatic fraction $\NCali/\NCaro$.

\begin{figure*}
\centerline{
\includegraphics[scale=0.55,clip]{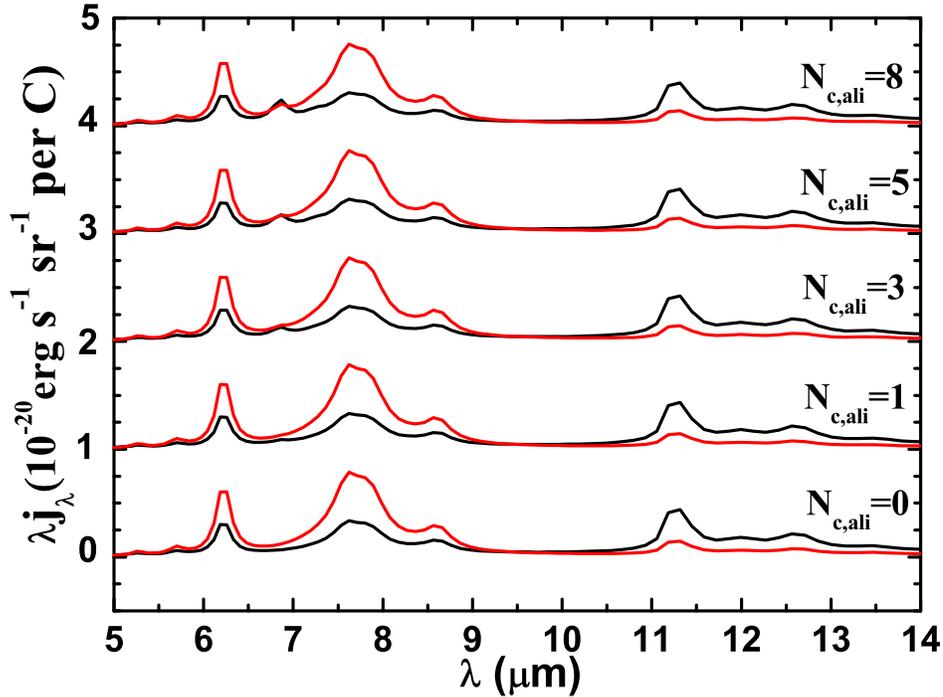}
}
\caption{\label{fig:irem}
         IR emission spectra of neutral (black lines)
         and ionized (red lines) methyl PAHs
         of $\NCali=0, 1, 3, 5, 8$ aliphatic C atoms
         and $\left(80-\NCali\right)$ aromatic C atoms
         illuminated by the MMP83 ISRF.
         For clarity, the spectra for methyl PAHs
         with $\NCali=1, 3, 5, 8$ are vertically shifted.
         }
\end{figure*}

\begin{figure*}
\centerline{
\includegraphics[scale=0.7,clip]{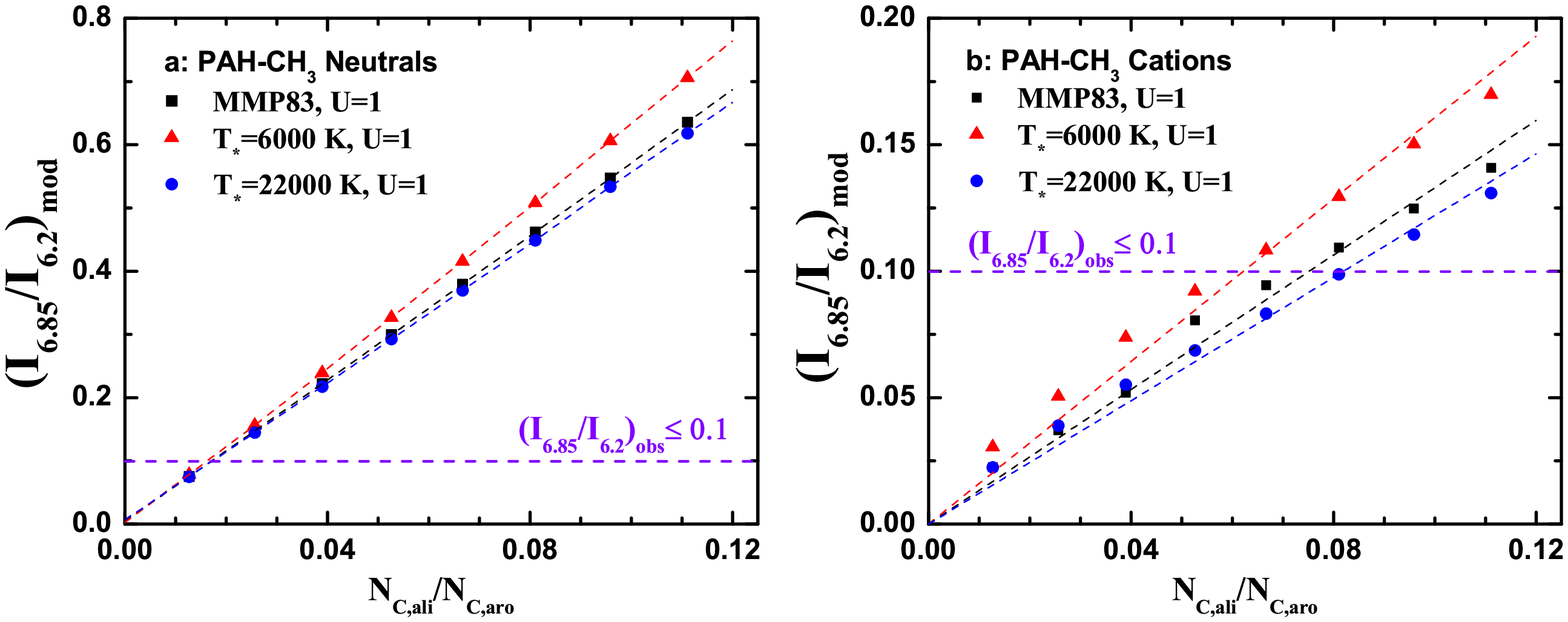}
}
\caption{\label{fig:I685I62vsAli2Aro}
              Model-calculated intensity ratios $\Iratiomod$
              as a function of the aliphatic fraction $\NCali/\NCaro$
              for (a) neutral methyl PAHs and (b) their cations.
              The molecules and their cations are illuminated by
              the MMP83 ISRF (squares),
              a solar-type star of $\Teff=6000\K$ (triangles),
              and a B1.5V star of $\Teff=22,000\K$ (circles).
              The dashed horizontal line plots the observed
              upper limit of the intensity ratio $\Iratioobs\simlt0.1$.
              The lines passing through the origin are linear fits
               to the model-calculated intensity-ratio data points.
              }
\end{figure*}

Let $dP$ be the probability that the temperature
of the methyl molecule will be in $[T,T+dT]$.
The emissivity of this molecule becomes
\begin{equation}
j_\lambda(\NC) = \int C_{\rm abs}(\NC,\lambda)\,
            4\pi B_\lambda(T)\,\frac{dP}{dT}\,dT  ~.
\end{equation}
As shown in Figures~6, 7 of Draine \& Li (2007),
the 6--8$\mum$ interstellar UIE emitters  are
in the size range of $\NC$\,$\simali$50--100 C atoms.
For illustrative purpose, we therefore consider $\NC=80$.
We adopt the ``thermal-discrete'' method of Draine \& Li (2007)
to compute the temperature probability distribution functions
of both neutral and ionized methyl PAHs of $\NCali=0, 1, 2, ...8$
aliphatic C atoms and $\left(80-\NCali\right)$
aromatic C atoms.
In Figure~\ref{fig:irem} we show the IR emission spectra of
both neutral and ionized methyl PAHs of $\NCali=0, 1, 3, 5, 8$
illuminated by the solar neighbourhood interstellar radiation
field (ISRF) of Mathis, Mezger \& Panagia (1983; MMP83).
Figure~\ref{fig:irem} shows that, while the 6.85$\mum$ feature
is clearly visible in the IR emission spectrum
for $\NCali=3$, the 7.25$\mum$ feature
remains hardly noticeable even for $\NCali=8$.
We will discuss this in more detail in \S\ref{sec:discussion}.
For a given $\NCali$, we derive $\Iratiomod$,
the model intensity ratio of the 6.85$\mum$ band to
the 6.2$\mum$ band, from
\begin{equation}\label{eq:Iratiomod}
\left(\frac{I_{6.85}}{I_{6.2}}\right)_{\rm mod}
= \frac{\int_{6.85}\Delta j_\lambda(\NC)\,d\lambda}
{\int_{6.2}\Delta j_\lambda(\NC)\,d\lambda} ~~,
\end{equation}
where $\int_{6.2}\Delta j_\lambda(\NC)\,d\lambda$
and $\int_{6.85}\Delta j_\lambda(\NC)\,d\lambda$
are respectively the feature-integrated excess emission
of the methyl PAH above the 6.2 and 6.85$\mum$ features.
In Figure~\ref{fig:I685I62vsAli2Aro} we show the model
intensity ratios $\Iratiomod$ as a function of $\NCali/\NCaro$.
For an observational upper limit of $\Iratioobs\simlt0.1$,
Figure~\ref{fig:I685I62vsAli2Aro} allows us to place an upper
limit of $\NCali/\NCaro\approx1.8\%$ for neutral methyl PAHs
and $\NCali/\NCaro\approx7.5\%$ for ionized methyl PAHs.
We therefore can also conclude that the carriers of
the 6--8$\mum$ UIE bands are largely aromatic,
with $<$\,10\% of the C atoms in aliphatic form.

\section{Discussion}\label{sec:discussion}

In \S\ref{sec:A685A62} we have shown that the aliphatic
fractions derived from the simple single-temperature
approach ($\NCali/\NCaro\approx0.96\%$ for neutrals
and $\NCali/\NCaro\approx9.6\%$ for cations)
are generally consistent with that derived from
the more rigorous calculations of the model emission
spectra ($\NCali/\NCaro\approx1.8\%$ for neutrals
and $\NCali/\NCaro\approx7.5\%$ for cations).
This can be understood in terms of the single-photon
heating process. Upon absorption of a photon of energy
$h\nu$ (where $\nu$ is the photon frequency),
a methyl PAH molecule will be heated to a maximum
temperature $T_{p}$ determined by its specific heat
$C(T)$ and $h\nu$:
$\int_{0}^{T_{p}} C(T)\,dT = h\nu$.
The molecule will then rapidly cool down and
radiate away most of the absorbed energy
at temperature $T_{p}$. Before it absorbs another
photon, the molecule will spend most of the time at
very low temperatures (Draine \& Li 2001).
Therefore, a single temperature of $T_{p}$
provides a reasonably good measure of the IR emission.

The aliphatic fractions derived
in \S\ref{sec:A685A62} from
the model intensity ratio $\Iratiomod$
%(see eq.\,\ref{eq:Iratiomod})
are calculated for methyl PAHs
excited by the MMP83 ISRF.
If the actual radiation intensity is $U$-times
that of the MMP83 ISRF, the results will remain
unchanged because the IR emission spectra of
stochastically heated nano-sized grains or
very large molecules, after scaled by $U$,
are essentially independent of $U$
(e.g., see Figure~13 of Li \& Draine 2001,
Figure~1f of Draine \& Li 2007).
To examine whether and how the ``hardness''
of the exciting starlight affects the aliphatic
fraction estimate, we consider methyl PAHs
of $\NCali=0, 1, 2, ...8$ aliphatic C atoms
and $\left(80-\NCali\right)$ aromatic C atoms
excited by stars with an effective temperature
of $\Teff=6,000\K$ like our Sun
and by stars of $\Teff=22,000\K$ like the B1.5V star
HD\,37903 which illuminates the reflection nebula NGC\,2023.
The starlight intensity in the 912$\Angstrom$--1$\mum$
wavelength range is fixed at $U=1$, with $U$ defined as
\begin{equation}
U \equiv \frac{\int_{1\mu {\rm m}}^{912{\rm \Angstrom}}
               4\pi J_\star(\lambda,\Teff)\,d\lambda}
              {\int_{1\mu {\rm m}}^{912{\rm \Angstrom}}
               4\pi J_{\rm ISRF}(\lambda)\,d\lambda} ~~,
\end{equation}
where $J_\star(\lambda, \Teff)$ is the intensity
of starlight approximated by the Kurucz model
atmospheric spectrum,
and $J_{\rm ISRF}(\lambda)$
is the MMP83 ISRF starlight intensity.
As shown in Figure~\ref{fig:I685I62vsAli2Aro},
for a given $\NCali/\NCaro$,
the $\Teff=6,000\K$ model results in
a larger $\Iratiomod$ than that of the MMP83 ISRF model,
while the $\Teff=22,000\K$ model results in
a smaller $\Iratiomod$ than that of the MMP83 ISRF model.
This is because a {\it softer} radiation field
rises the methyl PAH molecule to a {\it lower} temperature
and therefore it radiates relatively more at 6.82$\mum$
than at 6.2$\mum$.
Nevertheless, the effects on the derived $\NCali/\NCaro$
are small: the MMP83 ISRF model, the $\Teff=6,000\K$ model,
and the $\Teff=22,000\K$ model respectively derive
$\NCali/\NCaro\approx1.6\%, 1.8\%, 1.8\%$ for neutrals,
and $\NCali/\NCaro\approx6.2\%, 7.5\%, 8.2\%$ for cations
(see Figure~\ref{fig:I685I62vsAli2Aro}).

\begin{figure*}
\centerline{
\includegraphics[scale=0.25,clip]{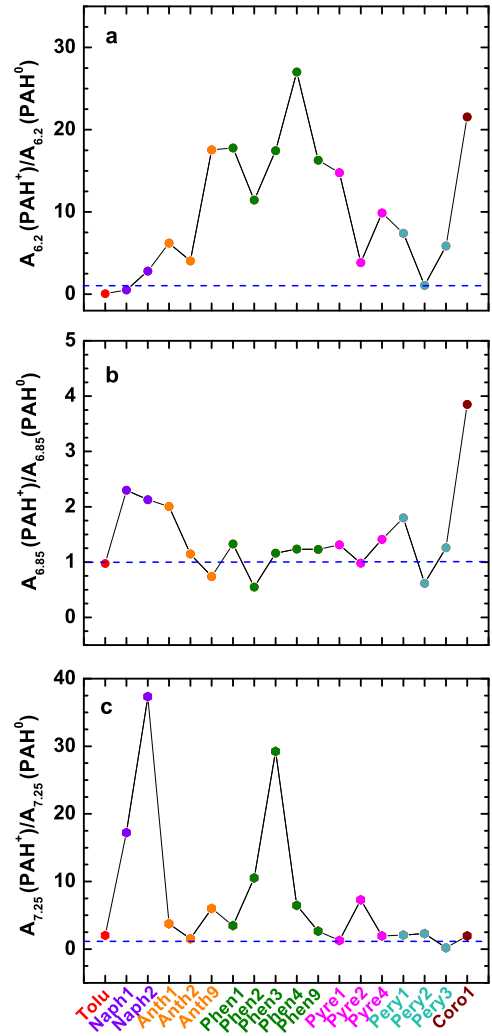}
}
\caption{\label{fig:Ion_Eff}
         Effects of ionization on
         the intrinsic strengths of
         the 6.2, 6.85 and 7.25$\mum$ bands.
         }
\end{figure*}

To examine the effects of ionization of
methyl PAHs on the intrinsic strengths of
the 6.2, 6.85 and 7.25$\mum$ bands,
we show in Figure~\ref{fig:Ion_Eff}
the intrinsic strengths of
these bands of ionized methyl PAHs
relative to that of their neutral counterparts.
As is long known (Allamandola et al.\ 1999,
Hudgins \& Allamandola 2005),
the 6.2$\mum$ C--C stretch
is substantially enhanced upon ionization
(see Figure~\ref{fig:Ion_Eff}a).
In contrast, the 6.85$\mum$ aliphatic C--H
deformation is only moderately enhanced
(see Figure~\ref{fig:Ion_Eff}b),
while the 7.25$\mum$ alipahtic C--H deformation
is more like the 6.2$\mum$ C--C stretch
and shows considerable enhancement upon ionization
(see Figure~\ref{fig:Ion_Eff}c).

\begin{figure*}
\centerline{
\includegraphics[scale=0.7,clip]{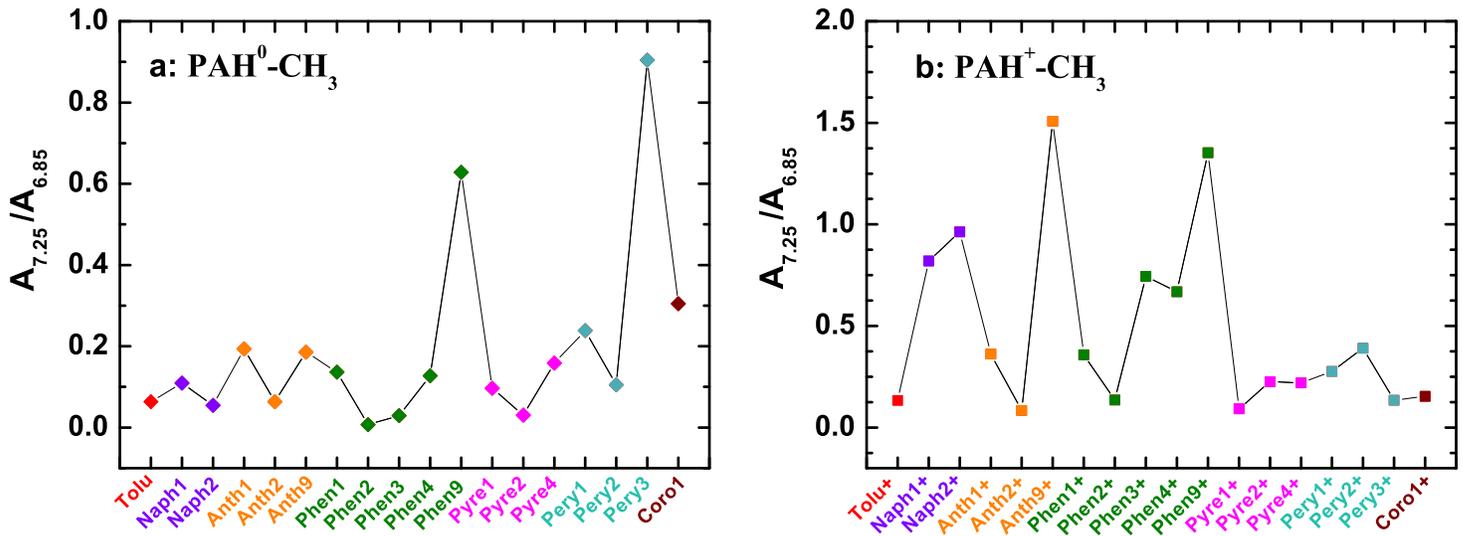}
}
\caption{\label{fig:A725A685}
         Comparison of the intrinsic strength
         of the 7.25$\mum$ aliphatic C--H deformation
         with that of the 6.85$\mum$ aliphatic
         C--H deformation for (a) neutral methyl PAHs
         and (b) ionized methyl PAHs.
         }
\end{figure*}

As mentioned in \S\ref{sec:obs} and \S\ref{sec:A685A62},
the detection of the 7.25$\mum$ emission feature
in the Milky Way or the Magellanic Clouds 
is rarer than the 6.85$\mum$ feature.
This is because, as shown in Figure~\ref{fig:A725A685},
the intrinsic strength of the 7.25$\mum$ feature is
weaker than that of the 6.85$\mum$ feature by a factor
of $\simali$8 for neutral methyl PAHs
and by a factor of $\simali$3 for their cations.
In addition, the weak 7.25$\mum$ feature could be
hidden by the pronounced 7.7$\mum$ C--C stretch.

The aliphatic fraction of the UIE carriers
derived from the 6.85$\mum$ aliphatic C--H
deformation is consistent with that derived
from the 3.4$\mum$ aliphatic C--H stretch
in the sense that the UIE carrier is
predominantly aromatic and the aliphatic component
is only a minor part of the UIE emitters.
Based on the observed intensity ratios of
the 3.4$\mum$ feature to the 3.3$\mum$ feature,
Li \& Draine (2012) placed an upper limit of
$\simali$9\% on the aliphatic fraction
of the UIE carriers in NGC\,7027
and the Orion bar.
Yang et al.\ (2013) further constrained
the aliphatic fraction to be at most $\simali$2\%
by examining a large sample of 35 UIE sources
which exhibit both the 3.3$\mum$
and 3.4$\mum$ C--H features.
Unlike Li \& Draine (2012) who adopted
the 3.3$\mum$ feature strength ($A_{3.3}$)
of small neutral PAHs (Draine \& Li 2007)
and the 3.4$\mum$ feature strength ($A_{3.4}$)
obtained by averaging over three pure aliphatic
molecules (i.e., ethane, hexane, and methyl-cyclo-hexane)
and one aromatic molecule with an aliphatic sidegroup
(i.e., ethyl-benzene; d'Hendecourt \& Allamandola 1986),
Yang et al.\ (2013) employed density functional theory
to compute $A_{3.4}/A_{3.3}$
for a range of methyl-substituted PAHs.

Li \& Draine (2012) have also examined
the 6.85 and 7.7$\mum$ features of
NGC\,7027 and the Orion bar
by comparing the observed intensity ratios
of these two bands ($I_{6.85}/I_{7.7}$)
with their intrinsic band strengths
($A_{6.85}$, $A_{7.7}$).
Adopting the 7.7$\mum$ feature strength
of charged aromatic molecules
(Draine \& Li 2007) for $A_{7.7}$
and the 6.85$\mum$ feature strength
obtained by averaging over that measured
for methylcylcohexane (d'Hendecourt \& Allamandola 1986)
and for HAC (Dartois \& Mu\~noz-Caro 2007) for $A_{6.85}$,
they derived the aliphatic fraction
of the UIE emitters to be $<$15\%.
With the band-strength ratio $\Adfa/\Acc$
obtained over many methylated aromatic species
(see \S\ref{sec:Method}),
we derive an aliphatic fraction of $<$10\%
from the observed intensity ratio $\Idfa/\Icc$
(see \S\ref{sec:A685A62}).

Finally, we note that the aliphatic C atoms
considered here are all in the form of methyl side groups.
The aliphatic fraction derived in \S\ref{sec:A685A62}
is the ratio between the number of C atoms
in the aromatic skeleton of PAHs and the number
of C atoms in methyl side groups.
This is not necessarily {\it all} of the aliphatic C atoms
in astronomical PAHs:
there may be some in linear bridges,
and there may also be some that are
in the regular PAH skeleton
but at a superhydrogenated site.\footnote{%
  To bond to an additional H atom,
  a C atom has to switch from sp$^2$ to sp$^3$ orbitals,
  thereby by definition becoming aliphatic.
  }
These aliphatic C atoms are not yet included in this study.
But Yang et al. (2016) have shown that the aliphatic fraction
of the UIE carriers derived from the $\Aratio$ ratios of
PAHs with a wide range of sidegroups (e.g., ethyl, propyl, butyl,
dimethyl) was close to that based on the $\Aratio$ ratios of
mono-methyl PAHs.
Also, the sample of regular and methylated PAHs
considered here (see Figure~\ref{fig:MonoMethylPAHs})
may not be representative of what are actually
present in astronomical environments
(e.g., the actual astro-PAH molecules may have
$\NC\simgt40$ C atoms [Tielens 2008]
while the largest molecule considered here
is coronene of $\NC=24$).
Future experimental and/or quantum-chemical computational
studies of larger species
would be very helpful to obtain more reliable estimates
of the $\Adfa/\Acc$ ratio, and therefore,
of the aliphatic fraction of the UIE carriers.
%
%

%%% More discussions: 3.3/3.4, 6.2/6.85um *absorption* %%%
%%% Why 3.4um is seen more often, while 6.85um rarely seen? %%%

\section{Summary}\label{sec:summary}
We have examined the nature of
the UIE emitters
by comparing the observed intensity ratios
of the 6.2$\mum$ aromatic C--C stretch
and the 6.85$\mum$ aliphatic C--H deformation
with the intrinsic band strengths of these two features
computed from density functional theory
for seven PAH molecules (and their cations)
with a methyl side chain.
We derive the fraction of C atoms
in methyl(ene) aliphatic form to be at most $\simali$10\%,
confirming the earlier finding that
the UIE emitters are largely aromatic.

\section*{Acknowledgements}
We thank T.~Onaka, J.Y.~Seok, G.~Sloan,
and the anonymous referee
for very helpful suggestions.
AL and XJY are supported in part by
NSFC\,11473023, NSF AST-1109039, NNX13AE63G,
Hunan Provincial NSF 2015JJ3124,
and the University of Missouri Research Board.
RG is supported in part by NSF-PRISM grant
Mathematics and Life Sciences (0928053).
Acknowledgment is also made to the donors of
the American Chemical Society Petroleum Research Fund
for partial support of this research (53415-ND4).
Computations were performed using the high-performance computer
resources of the University of Missouri Bioinformatics Consortium.

%%%%% Table 1 %%%%%
\begin{table*}
\footnotesize
\begin{center}
\caption[]{\footnotesize
           IR Intensity (km\,mol$^{-1}$) of
           the 6.85 and 7.25$\mum$ Aliphatic
           C--H Deformation Bands
           and the 6.2$\mum$ Aromatic C--C Stretch Band
           Computed at the {\rm B3LYP/6-311+G$^{\ast\ast}$} Level
           for All the Neutral Methyl PAHs
           Shown in Figure~\ref{fig:MonoMethylPAHs}.
           }
\label{tab:A685A62n}
\begin{tabular}{lccccc}
\noalign{\smallskip} \hline \hline \noalign{\smallskip}
Compound	&	$A_{6.2}$$^{a}$	
                &	$A_{6.85}$$^{b}$	
                &	$A_{6.85}/A_{6.2}$
                &	$A_{7.25}$$^{c}$	
                &	$A_{7.25}/A_{6.2}$
\\ \noalign{\smallskip} \hline \noalign{\smallskip}
Toluc	&	1.57 	&	6.92 	&	4.40 	&	0.44 	&	0.28 	\\
Naph1a	&	0.81 	&	5.25 	&	6.53 	&	0.58 	&	0.71 	\\
Naph2b	&	1.76 	&	4.63 	&	2.63 	&	0.25 	&	0.14 	\\
Anth1a	&	0.68 	&	6.83 	&	9.97 	&	1.32 	&	1.93 	\\
Anth2b	&	1.66 	&	7.82 	&	4.70 	&	0.49 	&	0.30 	\\
Anth9a	&	0.41 	&	8.66 	&	20.98 	&	1.61 	&	3.89 	\\
Phen1a	&	0.61 	&	9.94 	&	16.21 	&	1.36 	&	2.21 	\\
Phen2b	&	1.27 	&	11.04 	&	8.68 	&	0.08 	&	0.06 	\\
Phen3b	&	1.33 	&	8.05 	&	6.05 	&	0.24 	&	0.18 	\\
Phen4a	&	0.40 	&	7.84 	&	19.39 	&	1.00 	&	2.47 	\\
Phen9a	&	0.69 	&	4.05 	&	5.87 	&	2.54 	&	3.69 	\\
Pyre1a	&	0.82 	&	7.68 	&	9.42 	&	0.74 	&	0.86 	\\
Pyre2c	&	2.37 	&	6.28 	&	2.65 	&	0.19 	&	0.08 	\\
Pyre4a	&	1.23 	&	4.85 	&	3.94 	&	0.77 	&	0.59 	\\
Pery1c	&	1.26 	&	4.90 	&	3.89 	&	1.17 	&	0.93 	\\
Pery2a	&	3.01 	&	6.80 	&	2.26 	&	0.71 	&	0.24 	\\
Pery3a	&	1.83 	&	5.75 	&	3.14 	&	5.20 	&	2.84 	\\
Coro1a	&	0.79 	&	4.80 	&	6.09 	&	1.46 	&	1.86 	\\
\hline
\noalign{\smallskip} \noalign{\smallskip}
\end{tabular}
\begin{description}
\item[$^{a}$] Intensity of the aromatic C--C stretch
              at 6.2$\mum$ in $\km\mol^{-1}$ (per C--C bond).
\item[$^{b}$] Intensity of the aliphatic C--H deformation
              at 6.85$\mum$ in $\km\mol^{-1}$ (per C--H bond).
\item[$^{c}$] Intensity of the aliphatic C--H deformation
              at 7.25$\mum$ in $\km\mol^{-1}$ (per C--H bond).

\end{description}
\end{center}
\end{table*}
%%%%% Table 1 %%%%%

%%%%% Table 2 %%%%%
\begin{table*}
\footnotesize
\begin{center}
\caption[]{\footnotesize
           Same as Table~\ref{tab:A685A62n}
           but for the cations of all the isomers of
           the methyl PAHs
           shown in Figure~\ref{fig:MonoMethylPAHs}.
           }
\label{tab:A685A62i}
\begin{tabular}{lccccc}
\noalign{\smallskip} \hline \hline \noalign{\smallskip}
Compound	&	$A_{6.2}$
                &	$A_{6.85}$
                &	$A_{6.85}/A_{6.2}$
                &	$A_{7.25}$
                &	$A_{7.25}/A_{6.2}$
\\ \noalign{\smallskip} \hline \noalign{\smallskip}
Toluc+	&	0.10 	&	6.73 	&	66.27 	&	0.90 	&	8.82 	\\
Naph1a+	&	0.42 	&	12.08 	&	28.66 	&	9.91 	&	23.51 	\\
Naph2a+	&	4.92 	&	9.86 	&	2.00 	&	9.49 	&	1.93 	\\
Anth1a+	&	4.25 	&	13.70 	&	3.22 	&	4.95 	&	1.17 	\\
Anth2b+	&	6.74 	&	8.96 	&	1.33 	&	0.76 	&	0.11 	\\
Anth9b+	&	7.24 	&	6.40 	&	0.88 	&	9.66 	&	1.33 	\\
Phen1a+	&	10.90 	&	13.17 	&	1.21 	&	4.71 	&	0.43 	\\
Phen2b+	&	14.56 	&	6.04 	&	0.42 	&	0.82 	&	0.06 	\\
Phen3b+	&	23.22 	&	9.37 	&	0.40 	&	6.97 	&	0.30 	\\
Phen4a+	&	10.92 	&	9.68 	&	0.89 	&	6.46 	&	0.59 	\\
Phen9a+	&	11.24 	&	4.98 	&	0.44 	&	6.73 	&	0.60 	\\
Pyre1a+	&	12.70 	&	10.07 	&	0.79 	&	0.93 	&	0.07 	\\
Pyre2c+	&	9.59 	&	6.15 	&	0.64 	&	1.39 	&	0.14 	\\
Pyre4a+	&	12.83 	&	6.84 	&	0.53 	&	1.50 	&	0.12 	\\
Pery1c+	&	9.32 	&	8.81 	&	0.95 	&	2.43 	&	0.26 	\\
Pery2c+	&	3.21 	&	4.19 	&	1.30 	&	1.64 	&	0.51 	\\
Pery3a+	&	10.74 	&	7.24 	&	0.67 	&	0.98 	&	0.09 	\\
Coro1a+	&	16.98 	&	18.47 	&	1.09 	&	2.83 	&	0.17 	\\
\hline
\noalign{\smallskip} \noalign{\smallskip}
\end{tabular}
\end{center}
\end{table*}
%%%%% Table 2 %%%%%

%%%%% Table 3 %%%%%
\begin{table*}
\footnotesize
\begin{center}
\caption[]{\footnotesize
           Observed Intensity Ratios of
           the 6.85 and 7.25$\mum$
           Aliphatic C--H Deformation Bands
           to the 6.2$\mum$ Aromatic C--C Stretch Band
           ($I_{6.85}/I_{6.2}$, $I_{7.25}/I_{6.2}$).
           }
\label{tab:I685I62}
\begin{tabular}{lcccr}
\noalign{\smallskip} \hline \hline \noalign{\smallskip}
Object & Type
       & $I_{6.85}/I_{6.2}$
       & $I_{7.25}/I_{6.2}$
       &  Reference\\
\noalign{\smallskip} \hline \noalign{\smallskip}
NGC~7023 & RN & 0.043 & ... & This work\\
NGC~7027 & PN & 0.092 & ... & Kwok \& Zhang (2011)\\
Orion Bar & PDR & 0.097 & ... & Kwok \& Zhang (2011)\\
%\hline
IRAS~01005+7910	& PPN &	0.19 & ... & Zhang et al.\ (2010)\\
IRAS~04296+3429	& PPN &	1.30 & ... & Zhang et al.\ (2010)\\
IRAS~22223+4327	& PPN & 1.86 & ... & Zhang et al.\ (2010)\\
IRAS~22272+5435	& PPN & 4.23 & ... & Tokunaga (1997)\\
\hline
IRAS~00350-7436	& PPN
                & -0.72$\pm$0.13 & 0.47$\pm$0.24
                & Sloan et al.\ (2014)\\
IRAS~05063-6908	& PPN
                & 0.35$\pm$0.09 & 0.03$\pm$0.04
                & Sloan et al. (2014)\\
IRAS~05073-6752	& PPN
                & 1.05$\pm$0.15 & 0.00$\pm$0.04
                & Sloan et al.\ (2014)\\
IRAS~05092-7121	& PPN & 0.09 	$\pm{	0.01 	}$     & -0.02 	$\pm{	0.01 	}$     & Sloan et al. (2014)\\
IRAS~05110-6616	& PPN & 1.66 	$\pm{	0.06 	}$     & 0.34 	$\pm{	0.02 	}$     & Sloan et al. (2014)\\
IRAS~05185-6806	& PPN & 0.06 	$\pm{	0.01 	}$     & -0.02 	$\pm{	0.01 	}$     & Sloan et al. (2014)\\
IRAS~F05192-7008 & PPN & 0.23 	$\pm{	0.10 	}$     & 0.04 	$\pm{	0.09 	}$     & Sloan et al. (2014)\\
IRAS~Z05259-7052 & PPN & 0.33 	$\pm{	0.06 	}$     & 0.05 	$\pm{	0.01 	}$     & Sloan et al. (2014)\\
IRAS~05360-7121	& AGB & 0.05 	$\pm{	0.01 	}$     & -0.02 	$\pm{	0.01 	}$     & Sloan et al. (2014)\\
IRAS~05413-6934	& PPN & 0.27 	$\pm{	0.17 	}$     & -0.10 	$\pm{	0.10 	}$     & Sloan et al. (2014)\\
IRAS~05588-6944	& PPN & 0.14 	$\pm{	0.02 	}$     & -0.06 	$\pm{	0.02 	}$     & Sloan et al. (2014)\\
IRAS~06111-7023	& PPN & 1.05 	$\pm{	0.06 	}$     & 0.14 	$\pm{	0.01 	}$     & Sloan et al. (2014)\\
2MASS J00444111-7321361	& PPN & 1.03 	$\pm{	0.05 	}$     & 0.19 	$\pm{	0.02 	}$     & Sloan et al. (2014)\\
2MASS J01054645-7147053	& PPN & 0.09 	$\pm{	0.02 	}$     & 0.00 	$\pm{	0.03 	}$     & Sloan et al. (2014)\\
2MASS J05204385-6923403	& PPN & 0.42 	$\pm{	0.05 	}$     & 0.09 	$\pm{	0.03 	}$     & Sloan et al. (2014)\\
NGC 1978 WBT 2665 & PPN & 1.69 	$\pm{	0.32 	}$     & 0.28 	$\pm{	0.08 	}$     & Sloan et al. (2014)\\
SMP~SMC~011 & PN	& 0.06 	$\pm{	0.01 	}$     & -0.09 	$\pm{	0.02 	}$     & Sloan et al. (2014)\\
SMP~SMC~020 & PN  & 0.20 	$\pm{	0.08 	}$     & 0.14 	$\pm{	0.11 	}$     & Sloan et al. (2014)\\
\hline
\noalign{\smallskip} \noalign{\smallskip}
\end{tabular}
\end{center}
\end{table*}
%%%%% Table 3 %%%%%

%%%%%%%%%% Figure 1 %%%%%%%%%%%
%%%%%%%%%% Figure 1 %%%%%%%%%%%

%%%%%%%%%% Figure 2 %%%%%%%%%%%
%%%%%%%%%% Figure 2 %%%%%%%%%%%

%%%%%%%%%% Figure 3 %%%%%%%%%%%
%%%%%%%%%% Figure 3 %%%%%%%%%%%

%%%%%%%%%% Figure 4 %%%%%%%%%%%
%%%%%%%%%% Figure 4 %%%%%%%%%%%

%%%%%%%%%% Figure 5 %%%%%%%%%%%
%%%%%%%%%% Figure 5 %%%%%%%%%%%

%%%%%%%%%% Figure 6 %%%%%%%%%%%
%%%%%%%%%% Figure 6 %%%%%%%%%%%

%%%%%%%%%% Figure 7 %%%%%%%%%%%
%%%%%%%%%% Figure 7 %%%%%%%%%%%

%%%%%%%%%% Figure 8 %%%%%%%%%%%
%%%%%%%%%% Figure 8 %%%%%%%%%%%

%%%%%%%%%% Figure 9 %%%%%%%%%%%
%%%%%%%%%% Figure 9 %%%%%%%%%%%

%%%%%%%%%% Figure 10 %%%%%%%%%%%
%%%%%%%%%% Figure 10 %%%%%%%%%%%

\bsp
\label{lastpage}

\end{document}